\documentclass[a4paper,12pt,twoside,final]{article}

\usepackage[utf8]{inputenc}
\usepackage[in,plain]{fullpage}
\usepackage{amsmath,amssymb}
\usepackage{stix,bm}
\usepackage{cite}
\usepackage{microtype}

\usepackage[colorlinks,linkcolor=blue,citecolor=blue,urlcolor=blue]{hyperref}
\usepackage{graphicx}
\usepackage{tikz}
\usetikzlibrary{calc,patterns,decorations.pathmorphing,decorations.markings}

\usepackage[small,bf]{caption}

\makeatletter
\renewcommand\@makefntext[1]{\leftskip=0.0em\hskip-0.5em\@makefnmark#1}
\makeatother
\begin{document}

\renewcommand{\thefootnote}{\fnsymbol{footnote}}

\begin{titlepage}

\thispagestyle{empty}

\begin{center}

{\Large\textbf{Homogeneous model for the TRAPPIST-1e planet \\ with an icy layer}}

\vspace{8ex}

{\large\textbf{Yeva Gevorgyan}\footnote{Email: {\tt\href{mailto:yeva@ime.usp.br}{\nolinkurl{yeva@ime.usp.br}}}.}}

\centerline{\textit{Instituto de Matem\'{a}tica e Estat\'{i}stica, Universidade de S\~{a}o Paulo, 05508-090 S\~{a}o Paulo, SP, Brazil}}

\vspace{8ex}

{\large\textbf{Abstract}}

\vspace{2ex}

\parbox{36em}
{In this work we investigate whether a multilayered planet can be approximated as a homogeneous planet, and in particular how well the dissipation rate of a multilayered planet can be reproduced with a homogeneous rheology. We study the case of a stratified body with an icy crust that, according to recent studies, displays a double peak feature in the tidal response that cannot be reproduced with a homogeneous planet with an Andrade rheology. We revisit the problem with a slightly more complex rheology for the homogeneous body, the Sundberg-Cooper rheology, which naturally has a double peak feature, and apply the model to the TRAPPIST-1e planet. Our results compare very well with the results obtained when employing a multilayered model, showing that it is possible to approximate the behavior of a multilayer icy planet with a homogeneous planet using the Sundberg-Cooper rheology. This highlights the fact that we do not need the complexity of the multilayer planet model in order to estimate the tidal dissipation of an icy planet.

\vspace{4ex}

{\noindent}\textbf{Keywords}: \mbox{Terrestrial planets} $\cdot$ \mbox{planetary interiors} $\cdot$ \mbox{multi-layered planet} $\cdot$ \mbox{homogeneous body} ~$\cdot$ \mbox{Sundberg-Cooper rheology}}

\end{center}

\end{titlepage}


\section{Introduction}

Large amounts of data collected in recent decades from space missions as well as terrestrial and orbital observatories have allowed the structure and dynamics of celestial bodies in our Solar System to be constrained with ever increasing precision. Understanding the accurate observations of the rotation of the planets and icy moons of the Solar System currently requires, and allows for, the modeling of their stratified structures, which are typically composed of icy crusts, subsurface oceans, molten mantles, and solid cores, among other possibilities. Several models that account for the layered internal structure of these bodies have appeared in the recent literature \cite{2014Matsuyama,2018Matsuyama,HugoSylvio2017,Boue2017,Bolmont2020}. The more complex the model, the more parameters have to be fit from the data. For exoplanets and exomoons, the extremely limited information we have regarding their structure and dynamics severely constrains the number of parameters that can be employed in the models. In these cases, homogeneous models come in handy, with fewer parameters to fit. The question now is how well a multilayered planet can be approximated by a homogeneous body. The answer depends on the phenomena we are interested in.  Here we try to answer this question for the dissipation rate of a stratified body with an icy crust. The inspiration for this work comes from the recent paper by \cite{Bolmont2020}. In that paper the authors observe two peaks in the frequency dependence of the dissipation rate for a multilayered planet, TRAPPIST-1e, that has an icy crust. Here we study the same planet with parameters compatible with the ones in \cite{Bolmont2020} but modeled after a homogeneous rheology (Sundberg-Cooper) that is able to reproduce the double-peaked feature of the dissipation.

This paper is organized as follows. In Sect.~\ref{sec:double} we present the motivation of the work. In Sect.~\ref{sec:love} we present the equations required to calculate the dissipation rate for a homogeneous body with the Sundberg-Cooper rheology, which is a mixed Andrade-Maxwell-Voigt rheology, and in Sect.~\ref{sec:trapp} we apply the model to the motion of TRAPPIST-1e and compare the results  to those in \cite{Bolmont2020}. Section \ref{sec:conc} summarizes our results.


\section{The double-peaked feature of the dissipation for bodies with an ice layer}
\label{sec:double} 

In \cite{Bolmont2020} the authors apply their multilayered modeling to two types of planets: terrestrial planets and planets with an icy crust. For the terrestrial planets, the authors found that the tidal dissipation of the energy of stratified bodies is well approximated by that of a homogeneous model with the Andrade rheology. In the presence of an icy crust, the dissipation dependence on the frequency changes qualitatively at low frequencies: Two peaks in the dissipation appear; the first can be fit using the Andrade rheology, but the second requires a different rheology. At high frequencies, the Andrade rheology is once again applicable.

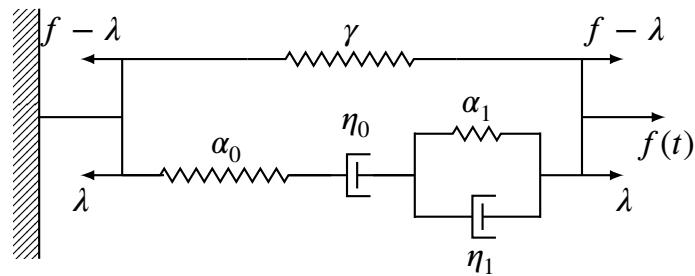
\begin{figure}
\begin{center}
\begin{tikzpicture}[scale=1.1, transform shape]
\tikzstyle{spring}=[thick,decorate,decoration={zigzag,pre length=0.5cm,post length=0.5cm,segment length=6}]
\tikzstyle{damper}=[thick,decoration={markings,
  mark connection node=dmp,
  mark=at position 0.5 with
  {
    \node (dmp) [thick,inner sep=0pt,transform shape,rotate=-90,minimum width=15pt,minimum height=3pt,draw=none] {};
    \draw [thick] ($(dmp.north east)+(5pt,0)$) -- (dmp.south east) -- (dmp.south west) -- ($(dmp.north west)+(5pt,0)$);
    \draw [thick] ($(dmp.north)+(0,-5pt)$) -- ($(dmp.north)+(0,5pt)$);
  }
}, decorate]
\tikzstyle{ground}=[fill,pattern=north east lines,draw=none,minimum width=0.75cm,minimum height=0.3cm]

            \draw [thick] (0,0.7) -- (1,0.7);
            \draw [thick] (1,0) -- (1,1.4);
            \draw [-latex, thick] (2.5,1.4) -- (0.5,1.4) node[above] {$f-\lambda$};
            \draw [-latex, thick] (1.5,0) -- (0.5,0) node[below] {$\lambda$};
            \draw [thick] (4.5,-0.515) -- (4.5,0.515);
            \draw [thick] (4,0) -- (4.5,0);

           \draw [spring] (1,0) -- node[above] {$\alpha_0$} (3.5,0);
            \draw [damper] (3,0) -- (4.5,0);
            \node at (3.8,0.6) {$\eta_{0}$};

            \draw [spring] (4.5,0.5) -- node[above] {$\alpha_1$} (6,0.5);
            \draw [damper] (4.5,-0.5) -- (6,-0.5);
            \node at (5.3,-1.05) {$\eta_{1}$};

            \draw [-latex,thick] (6,0) -- (7,0) node[below] {$\lambda$};
            \draw [thick] (6,-0.515) -- (6,0.515);
            \draw [spring] (2.5,1.4) -- node[above] {$\gamma$} (5,1.4);
            \draw [-latex, thick] (5,1.4) -- (7,1.4) node[above] {$f-\lambda$};
            \draw [thick] (6.5,0) -- (6.5,1.4);
            \draw [-latex, thick] (6.5,0.7) -- (7.5,0.7) node[below] {$f(t)$};

\node (wall) at (-0.15,0.5) [ground, rotate=-90, minimum width=3cm] {};
\draw [thick] (wall.north east) -- (wall.north west);
      \end{tikzpicture}
\end{center}
\caption[General oscillator with rheology]{General oscillator with a Burgers rheology. The external force $f(t)$ splits into the force $\lambda(t)$, which acts upon the rheology array, and the force $f(t)-\lambda(t),$ which acts upon the $\gamma$ spring.}
\label{burg-osc}
\end{figure}

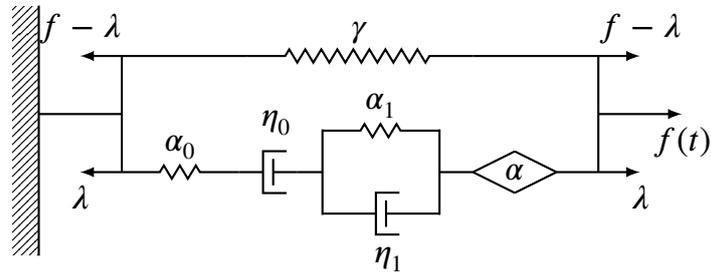
\begin{figure}
\begin{center}
\begin{tikzpicture}[scale=1.1, transform shape]
\tikzstyle{spring}=[thick,decorate,decoration={zigzag,pre length=0.5cm,post length=0.5cm,segment length=6}]
\tikzstyle{damper}=[thick,decoration={markings,
  mark connection node=dmp,
  mark=at position 0.5 with
  {
    \node (dmp) [thick,inner sep=0pt,transform shape,rotate=-90,minimum width=15pt,minimum height=3pt,draw=none] {};
    \draw [thick] ($(dmp.north east)+(5pt,0)$) -- (dmp.south east) -- (dmp.south west) -- ($(dmp.north west)+(5pt,0)$);
    \draw [thick] ($(dmp.north)+(0,-5pt)$) -- ($(dmp.north)+(0,5pt)$);
  }
}, decorate]
\tikzstyle{ground}=[fill,pattern=north east lines,draw=none,minimum width=0.75cm,minimum height=0.3cm]

            \draw [thick] (0,0.7) -- (1,0.7);
            \draw [thick] (1,0) -- (1,1.4);
            \draw [-latex, thick] (2.5,1.4) -- (0.5,1.4) node[above] {$f-\lambda$};
            \draw [-latex, thick] (1,0) -- (0.5,0) node[below] {$\lambda$};

            \draw [spring] (1,0) -- node[above] {$\alpha_0$} (2.4,0);
            \draw [damper] (2.4,0) -- (3.1,0);
            \node at (2.85,0.6) {$\eta_{0}$};

            \draw [thick] (3.1,0) -- (3.4,0);
            
            \draw [thick] (3.4,-0.515) -- (3.4,0.515);
            \draw [spring] (3.4,0.5) -- node[above] {$\alpha_1$} (4.8,0.5);
            \draw [damper] (3.4,-0.5) -- (4.8,-0.5);
            \node at (4.2,-1.05) {$\eta_{1}$};
            \draw [thick] (4.8,-0.515) -- (4.8,0.515);

            \draw [thick] (4.8,0) -- (5.2,0);

            \draw [thick] (5.2,0) -- (5.7,0.25);
            \draw [thick] (5.7,0.25) -- (6.2,0);
            \draw [thick] (6.2,0) -- (5.7,-0.25);
            \draw [thick] (5.7,-0.25) -- (5.2,0);
            \node at (5.7,0) {$\alpha$};

            \draw [-latex,thick] (6.2,0) -- (7.2,0) node[below] {$\lambda$};
            \draw [spring] (2.5,1.4) -- node[above] {$\gamma$} (5.2,1.4);
            \draw [-latex, thick] (5.2,1.4) -- (7.2,1.4) node[above] {$f-\lambda$};
            \draw [thick] (6.7,0) -- (6.7,1.4);
            \draw [-latex, thick] (6.7,0.7) -- (7.7,0.7) node[below] {$f(t)$};

\node (wall) at (-0.15,0.5) [ground, rotate=-90, minimum width=3cm] {};
\draw [thick] (wall.north east) -- (wall.north west);
      \end{tikzpicture}
\end{center}
\caption[General oscillator with rheology]{Same as Fig.~\ref{burg-osc} but for the Sundberg-Cooper rheology.}
\label{burgand-osc}
\end{figure}

 The simplest rheological model with two dissipation peaks at low frequencies is the well-known Burgers rheology in Fig.~\ref{burg-osc} \cite{Shoji2013}. It would then be natural to use a model that displays Burgers behavior at low frequencies and Andrade behavior at high frequencies. Such a composite viscoelastic model was proposed by \cite{Marshall2010} to incorporate both the effects of elastically accommodated grain boundary sliding and the transient diffusion creep observed in fine-grained peridotite. The viscoelastic oscillator with Sundberg-Cooper rheology can be seen in Fig.~\ref{burgand-osc}. The rheology was previously used to model the interiors of Mars, Io, and some exoplanets by \cite{Renaud2018} and \cite{Bagheri2019}.

\begin{figure}
\centering
        \includegraphics[width=0.75\columnwidth]{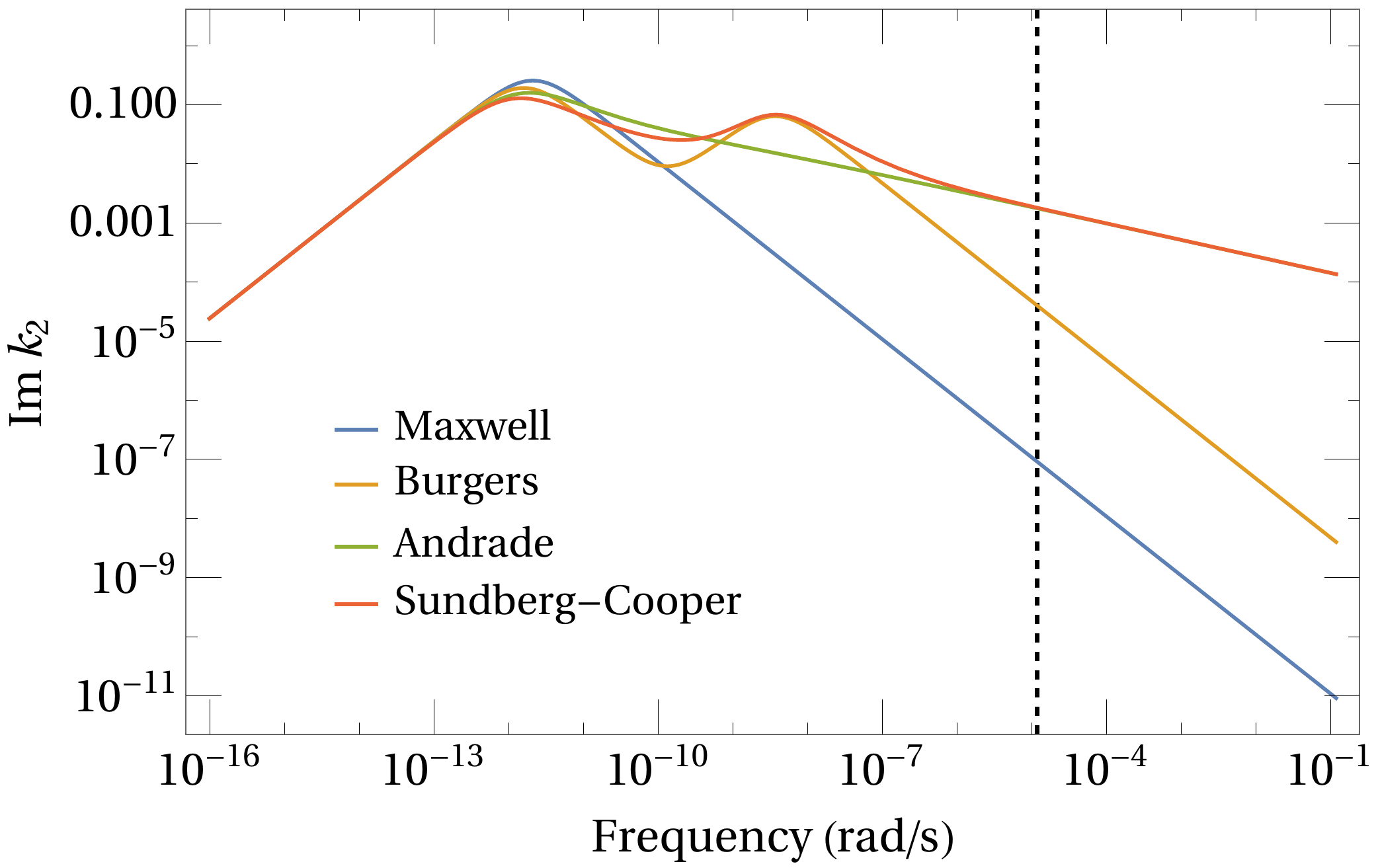}
    \caption{Imaginary part of the Love number for four different rheologies. The dashed vertical line marks the value of the TRAPPIST-1e mean rotation rate.}
    \label{fig:love_compare}
\end{figure}

The qualitative comparison of the dissipation rate for various rheologies can be seen in Fig.~\ref{fig:love_compare}. One can observe that the Sundberg-Cooper rheology indeed has two peaks in the dissipation rate at intermediate frequencies and preserves the Andrade behavior at high frequencies. The parameter values used to plot the figure are given in Tables~\ref{tab:parameters} and \ref{tab:trappist1e}. The details on how to calculate the energy dissipation rate with the Sundberg-Cooper rheology are discussed in the next section.

\begin{table}
\centering
\caption{Parameters for the viscoelastic elements in Fig.~\ref{burgand-osc}.}
\label{tab:parameters}
\begin{tabular}{llll} 
\hline
Quantity & Symb. & Unit & Value \\
\hline
Elasticity of Maxwell elem. & $\alpha_0$ & $s^{-2}$ & $5.815\times 10^{-7}$ \\
Viscosity of Maxwell elem. & $\eta_0$ & $s^{-1}$ & $181311$ \\
Elasticity of Voigt elem. & $\alpha_1$ & $s^{-2}$ & $2\times \alpha_0$ \\ 
Viscosity of Voigt elem. & $\eta_{1}$ & $s^{-1}$ & $4.15\times 10^{0}$--$10^{4}$ \\
Andrade parameter & $\alpha$ &  & $0.28$ \\
\hline
\end{tabular}
\end{table}
   
\begin{table}
\centering
\caption{Parameters for the viscoelastic elements used to obtain the values in Table~\ref{tab:parameters}. These parameters can be directly compared to the ones in \cite{Bolmont2020}.}
\label{tab:parameters2}
\begin{tabular}{lll} 
\hline
Quantity & Unit & Value \\
\hline
Elasticity of Maxwell elem. & Pa & $1.5\times9.333\times 10^{9}$ \\
Viscosity of Maxwell elem.  & Pa\,s & $4.365\times 10^{21}$ \\
Elasticity of Voigt elem.  & Pa & $3\times 9.333\times 10^{9}$ \\ 
Viscosity of Voigt elem.  & Pa\,s & $10^{17}$--$10^{21}$ \\
\hline
\end{tabular}
\end{table}

\section{Energy dissipation and the Love number}
\label{sec:love} 

In \cite{Gev2020}, the authors obtained an analytic formula for the tidal energy dissipation rate of a body in $1{:}1$ spin-orbit resonance on a slightly eccentric planar orbit in the presence of forced librations. If we discard the contribution from the forced librations (here we are not interested in this contribution, which could be added to the problem if needed), the equation for the average energy dissipation rate becomes
\begin{equation}\label{disen}
\frac{\Delta E}{T}=\; - \Im\left[k_{2}(\omega)\right]\frac{(nR)^5}{G}\left(\frac{m_2}{m_1+m_2}\right)^2\times\frac{21}{2}e^2,
\end{equation}
where $k_{2}(\omega)$, $\omega$, $n$, $R$, $G$, $m_1$, $m_2$, and $e$ are, respectively, the Love number, rotation frequency, mean rotation rate, mean radius, gravitational constant, mass of the planet, mass of the host body, and orbit eccentricity. 

The frequency-dependent part of the dissipation rate is encoded in the imaginary part of the Love number, $k_{2}(\omega)$. The following expression for $k_{2}(\omega)$ was obtained in \cite{CRR2018} using the ``association principle'' in the frequency domain,
\begin{equation}
\label{eq:K2(omega)}
k_{2}(\omega) = \frac{3G\rm{I}_{o}}{R^5}\frac{1}{\gamma+\hat{J}^{-1}(\omega)-\mu\omega^2},
\end{equation}
where $\gamma$, $\mu$, and $\hat{J}(\omega)$ are, respectively, the rigidity constant, inertia constant, and creep function, and where $\rm{I}_o$ is the mean moment of inertia of the planet. This formula is valid for an arbitrary rheology represented by a general oscillator.

The association principle was formulated by \cite{RR2017} and can be used to obtain equations of motion and rheology parameters, such as the Love number in the frequency domain, for an arbitrary rheology represented by a general oscillator, such as those in Figs.~\ref{burg-osc} and \ref{burgand-osc}. The one-dimensional general oscillators are constructed from spring-dashpot systems (bottom part in Figs.~\ref{burg-osc} and \ref{burgand-osc}), used to represent linear viscoelastic rheologies, by adding a parallel spring of elastic constant, $\gamma$, which represents gravity.

If the inertia of deformation is neglected ($\mu=0$), as assumed in this paper, then
\begin{equation}
\label{K2}
\begin{split}
  k_{2}(\omega)&=\frac{3G\rm{I}_{o}}{R^5}\frac{1}{\gamma+\hat{J}^{\,-1}(\omega)} \\
&=\frac{3G\rm{I}_{o}}{R^5}\frac{\gamma(\Re[\hat{J}(\omega)]^2+\Im[\hat{J}(\omega)]^2)+\Re[\hat{J}(\omega)]+i\Im[\hat{J}(\omega)]}{(\gamma \Re[\hat{J}(\omega)]+1)^2+\gamma^2 \Im[\hat{J}(\omega)]^2}.
\end{split}
\end{equation}
This expression is equivalent to the $k_{2}(n) = \frac{3}{2}\Big(1+\frac{57}{8\pi}\frac{1}{G\rho^2R^2}\frac{1}{\hat{J}}\Big)^{-1}$ used in \cite[Eq.~36]{Efroimsky2015}. A detailed comparison of the two notations can be found in \cite[Sect.~4]{CRR2018}. The imaginary part of the Love number that controls the behavior of the dissipation is given by
\begin{equation}\label{imk2}
\Im[k_{2}(\omega)]=\frac{3G\rm{I}_o}{R^5}\frac{\Im[\hat{J}(\omega)]}{(\gamma \Re[\hat{J}(\omega)]+1)^2+\gamma^2 \Im[\hat{J}(\omega)]^2}.
\end{equation}
For the Sundberg-Cooper rheology in Fig.~\ref{burgand-osc},
\begin{equation}
\hat{J}(\omega)=J_{0}\Bigg(1-i(\omega\tau_{M})^{-1}+\frac{J_1/J_{0}}{1+i\eta_{1} J\omega}+(i\omega\tau_{A})^{-\alpha}\Gamma(1+\alpha)\Bigg),
\end{equation}
where $J_0=1/\alpha_0$, $J_1=1/\alpha_1$, $\tau_{M}=\eta_0/\alpha_0$ is the Maxwell time and $\tau_{A}=(J_0/A)^{1/\alpha}$ is the Andrade time, with $\alpha$ and $A$ being parameters of the Andrade element. The other parameters are defined in Table~\ref{tab:parameters}. The physical dimensions of $\eta_{0}, \eta_{1}$ and $\alpha_{0}, \alpha_{1}$ are $1/s$ and $1/s^{2}$, respectively. These constants can be obtained from the parameters in Table~\ref{tab:parameters2} with the usual dimensions (viscosity[Pa s] and elasticity[Pa]) if the parameters are multiplied by 
\[
\frac{152\pi}{15}\frac{R}{m_{1}}=4.15375\times10^{-17} \rm{m/kg}.
\] 
The unusual dimension of these parameters and of $\gamma$ as well as the equivalence of the two sets of parameters are explained in detail in \cite[Sects.~2.1 and 4]{CRR2018}. Consequently,
\begin{equation}\label{eq:imj}
\Im[\hat{J}(\omega)]=-J_{0}\Bigg((\omega\tau_{M})^{-1} + \frac{\eta_{1}\omega J_1^2/J_{0}}{1+(\eta_{1} J_1\omega)^2} + \sin \frac{\pi\alpha}{2}(\omega\tau_{A})^{-\alpha}\Gamma(1+\alpha)\Bigg),
\end{equation}
and
\begin{equation}\label{eq:rej}
\Re[\hat{J}(\omega)]=-J_{0}\Bigg(1 + \frac{J_1/J_{0}}{1+(\eta_{1} J_1\omega)^2} + \cos \frac{\pi\alpha}{2}(\omega\tau_{A})^{-\alpha}\Gamma(1+\alpha)\Bigg).
\end{equation}
Equations (\ref{disen}), (\ref{imk2}), (\ref{eq:imj}), and (\ref{eq:rej}) combined are used in the next section to evaluate the dissipative behavior of the TRAPPIST-1e planet modeled after the Sundberg-Cooper rheology.


\section{TRAPPIST-1e}
\label{sec:trapp}

As an application of the Sundberg-Cooper rheology, we study the revolution of the TRAPPIST-1e planet. TRAPPIST-1e is a solid, close-to-Earth-sized exoplanet orbiting within the habitable zone around the ultra-cool dwarf star TRAPPIST-1 \cite{TRAPPIST2017,TRAPPIST2018}. We assume the planet to be a homogeneous body with a Sundberg-Cooper rheology that is experiencing tidal deformations due to its host star. The parameters and assumptions on the planet orbit are either the same as or close to the ones in \cite{Bolmont2020} and can be seen in Table~\ref{tab:trappist1e}. The mean moment of inertia and the rigidity constant in Table~\ref{tab:trappist1e} are estimated using the approach in \cite{CRR2018}. We assume that the planet is in $1{:}1$ spin-orbit resonance with its host red dwarf. The orbit is assumed to be planar and just slightly eccentric. 

\begin{figure}
\raggedleft
        \includegraphics[width=0.75\columnwidth]{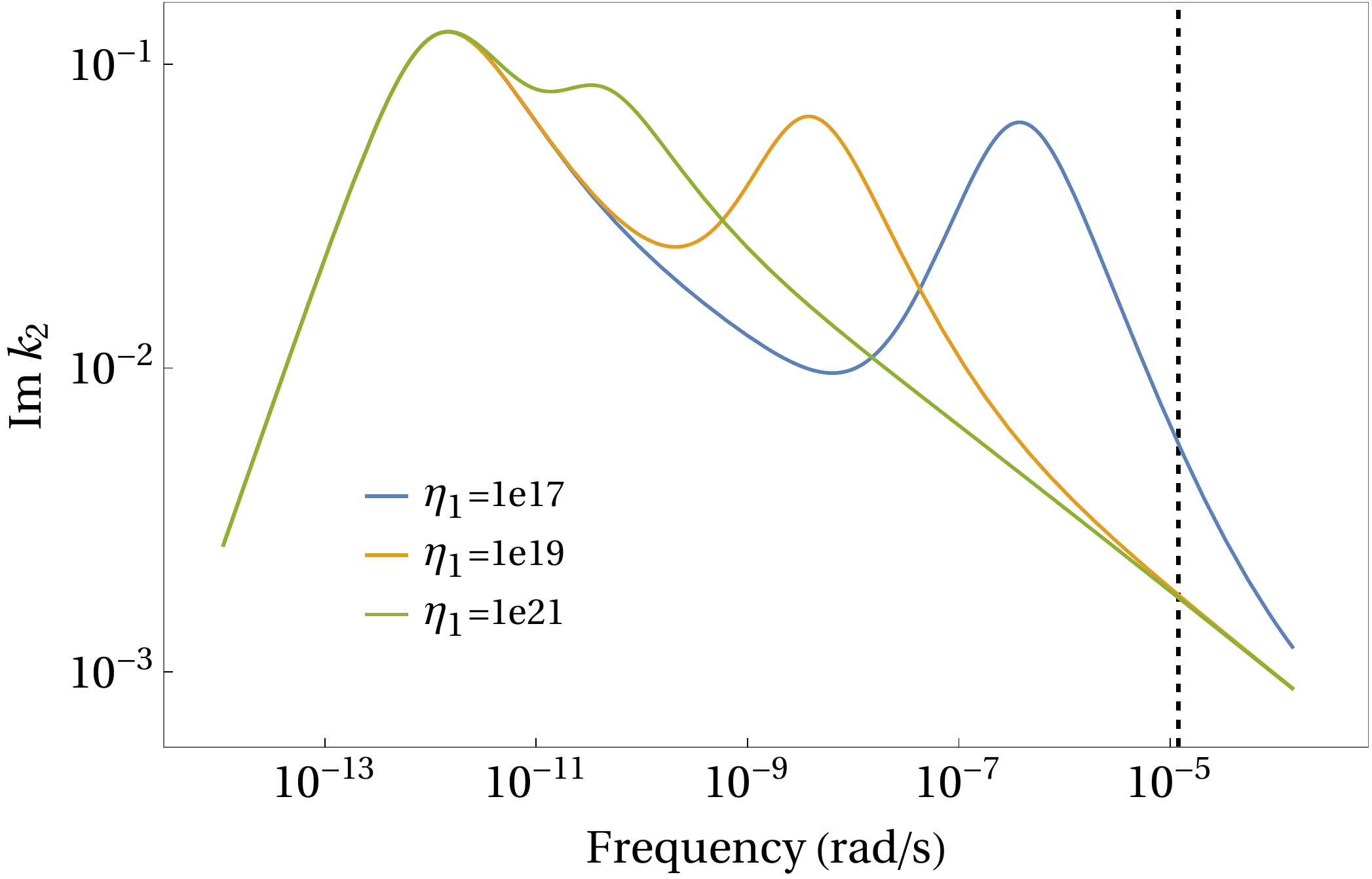}\\
        \includegraphics[scale=0.46]{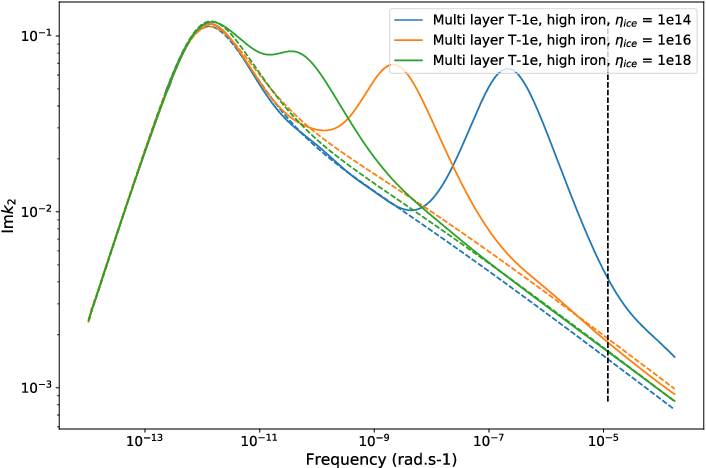}
    \caption{Comparison of dissipation for homogeneous and stratified rheologies. Top:~Imaginary part of the Love number for TRAPPIST-1e modeled after the Sundberg-Cooper rheology for different values of $\eta_{1}$ and $\tau_{A}=1.18\times\tau_{M}$. The dashed vertical line marks the value of the TRAPPIST-1e mean rotation rate. Bottom:~Same as top but for multilayered rheology. The figure is taken from \cite[Fig.~11]{Bolmont2020}.}
    \label{fig:eta_var}
\end{figure}

We are interested in reproducing the dissipation rate of the planet modeled after a multilayered rheology with an icy layer in \cite{Bolmont2020}. As mentioned before, it suffices to study the frequency dependence of the imaginary part of the Love number. In Fig.~\ref{fig:eta_var} (top) we plot $\Im[k_{2}(\omega)]$ against $\omega$ for three different values of the viscosity parameter, $\eta_{1}$, of the Voigt element ($10^{17}$, $10^{19}$, and $10^{21}$ Pa\,s) and observe the behavior of the second dissipation peak. We see that with increased viscosity the peak migrates to the left, closer and closer to the first peak. This behavior mimics the behavior in Fig.~\ref{fig:eta_var} (bottom), which was obtained with a multilayered model. The viscosities used in our model are $10^3$ times larger than those of the ice employed in \cite{Bolmont2020}. It is important to note that the viscosity of the Voigt element in our model is just part of a whole rheological model and, unlike in \cite{Bolmont2020}, does not relate directly to the actual ice crust in TRAPPIST-1e conditions. The remaining parameters were chosen such that we observe the same dissipation rate as that in \cite{Bolmont2020}; they are given in Table~\ref{tab:parameters}. Overall, one can see from Fig.~\ref{fig:eta_var} that the imaginary part of the Love number for the homogeneous model with the Sundberg-Cooper rheology (top) can reproduce the same curve for the multilayered body (bottom).

\begin{figure}
\centering
        \includegraphics[width=0.75\columnwidth]{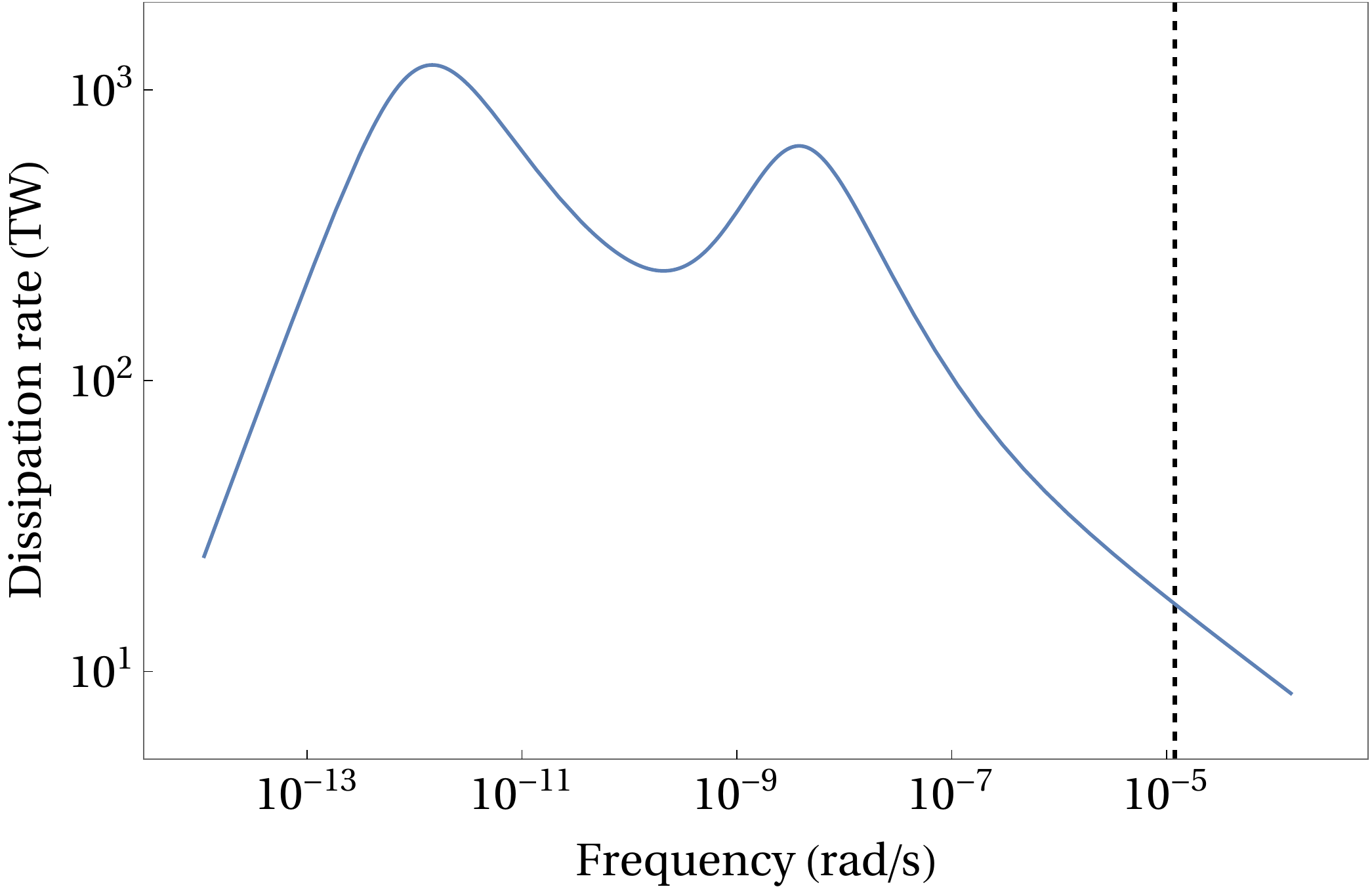}
    \caption{Dissipation rate for $\eta_{1}=10^{19}$~Pa\,s and $\tau_{A}=1.18\times\tau_{M}$.} The dashed vertical line marks the value of the TRAPPIST-1e mean rotation rate.
    \label{fig:dissipation}
\end{figure}

In Fig.~\ref{fig:dissipation} we plot the energy dissipation rate dependence on the frequency for $\eta_{1}=10^{19}$ Pa\,s and $e=0.00452$ using Eq.~(\ref{disen}). We estimate the dissipation rate for three possible orbit eccentricities and compare the results to the ones obtained by \cite{Bolmont2020} for a multilayered body with no icy layer. According to these authors, the dissipation rate for TRAPPIST-1e increases by ${\sim}15\%$ when an icy layer is assumed. One can see from Table~\ref{tab:dissipation} that the ${\sim}15\%$ difference is preserved for all three eccentricities.

\begin{table}
        \centering
        \caption{Orbital parameters for the TRAPPIST-1e planet taken from \cite{Bolmont2020} or calculated.}
        \label{tab:trappist1e}
        \begin{tabular}{llll} 
                \hline
                Quantity & Symb. & Unit & Value  \\
                \hline
                Mass & $m_{1}$ & kg & $4.6\times 10^{24}$  \\
                Mass of the star & $m_{2}$ & kg & $0.1796\times 10^{30}$  \\
                Radius & $R$ & km & 6002 \\
                Rotation period & $n$ & rad $s^{-1}$ & $1.19217\times 10^{-5}$   \\
                Semi-major axis & $a$ & AU & $0.029$ \\
                Mean moment of inertia & $\rm{I}_o$ & kg $m^2$ & $6.6284\times10^{37}$ \\
                Rigidity constant & $\gamma$ & $s^{-2}$ & $1.13592\times10^{-6}$\\
                \hline
        \end{tabular}
\end{table}

\begin{table}
\centering
\caption{Energy dissipation rate for three possible TRAPPIST-1e orbit eccentricities for the multilayer model (MLM) with no icy layer (or a low iron composition) taken from \cite{Bolmont2020} and for the homogeneous model (HM) with an icy layer obtained here.}
\label{tab:dissipation}
\begin{tabular}{lll} 
\hline
Eccentricity & MLM, no ice (TW) & HM, icy (TW) \\
\hline
$0.00452$ &  $9.3$ & $10.8$ \\
$0.00510$ & $12$   & $13.8$ \\
$0.00568$ & $15$   & $17.1$ \\ 
\hline
\end{tabular}
\end{table}

\section{Conclusions}
\label{sec:conc}
We have shown that a multilayered body with an icy layer can be reasonably approximated by a homogeneous body with the Sundberg-Cooper rheology as long as we are interested in the tidal dissipation of energy in the body. The mean dissipation rate at the rotation frequency as well as its behavior for the entire range of frequencies obtained here for the body modeled with an effective rheology are in very good agreement with the values obtained for the multilayered body in \cite{Bolmont2020}. Our results support the idea that for some bodies and ranges of parameters we do not really need to use complex multilayer models and that effective homogeneous models can be a good approximation for possibly stratified planets and satellites. It would be interesting to study the connection between the parameters of homogeneous and stratified models, such as the relationship between the viscosity parameters in the Sundberg-Cooper rheology and the viscosities of the actual components of the stratified bodies being modeled.

\section*{Acknowledgments}
The author thanks Clodoaldo G. Ragazzo (IME/USP) for useful discussions and a careful reading of the manuscript, and J.~Ricardo G. Mendonça (EACH/USP) and an anonymous referee for several remarks and many suggestions improving the manuscript. The author is partially supported by FAPESP, Brazil under grant no.~2019/25356-9.


\vspace{4ex}

\hspace*{\fill} $\star \quad \star \quad \star$ \hspace*{\fill}

\end{document}